\documentclass[reprint,amsmath,amssymb,aps]{revtex4-2}
\usepackage[utf8]{inputenc} 
\usepackage[T1]{fontenc}    
\usepackage[pdfusetitle]{hyperref}
\usepackage{graphicx,url,qrcode,bm,tikz,physics,dsfont,slashed}
\DeclareMathOperator\arctanh{arctanh}
\usetikzlibrary{calc}
\usepackage{cleveref}
\crefname{section}{Sec.}{Secs.}
\Crefname{section}{Section}{Sections}
\crefname{figure}{Fig.}{Figs.}
\Crefname{figure}{Figure}{Figures}
\crefname{equation}{Eq.}{Eqs.}
\crefname{appendix}{Appendix}{Appendices}
\newcommand{\im}[0]{\mathrm{i}}
\newcommand{\eu}[0]{\mathrm{e}}
\renewcommand{\vec}[1]{\bm{#1}}
\renewcommand{\Tr}[1]{\operatorname{Tr}\left\{#1\right\}}
\newcommand{\Heavi}[0]{\Theta }
\newcommand{\id}[0]{\mathds 1}
\newcommand{\diagram}[1]{\vcenter{\hbox{\begin{tikzpicture} #1 \end{tikzpicture}}}}
\usepackage[normalem]{ulem}

\newcommand{\dThree}[0]{2.7cm}
\newcommand{\dFour}[0]{3.6cm}
\newcommand{\dFive}[0]{4.5cm}

\newcommand{\len}[0]{1.0}

\tikzset{
  gluon/.style={thick, decorate, decoration={coil, amplitude=.08cm, segment length=.15cm, aspect=.75}}}
\tikzset{pion/.style={
	dashed,
	postaction={decorate},
	decoration={markings,mark=at position .6 with {\arrow{>}}}
	}}
\tikzset{nucleon/.style={
	postaction={decorate},
	decoration={markings,mark=at position .6 with {\arrow{>}}}
	}}
\tikzset{inmedium/.style={
	very thick,
	}}
\tikzset{source/.style={
	decorate,
	decoration={snake,amplitude=0.8mm}
	}}
\newcommand{\vertex}[2]{
	\fill[black!5] (#1) circle (.2);
	\draw (#1) circle (.2);
	\node at (#1) {\scalebox{0.8}{$#2$}};
}

\tikzset{>=latex}
\usetikzlibrary{datavisualization}
\usetikzlibrary{datavisualization.formats.functions}
\usetikzlibrary{patterns}    	             
\usetikzlibrary{decorations.markings}        
\usetikzlibrary{shapes.geometric}            
\usetikzlibrary{decorations.pathmorphing}    
\usetikzlibrary{decorations}

\newcommand{\CondensateTypeTwoPi}[0]{
  \coordinate (A) at (0,0);
  \coordinate (B) at ($(A)+(\len,0)$);
  \coordinate (C) at ($(B)+(\len,0)$);
  \coordinate (C2) at ($(C)+(\len,0)$);
  \coordinate (D) at ($(B)+(0,-0.6)$);
  \draw[source] (A) -- (B);
  \draw[pion] (B) -- (C);
  \draw[source] (C) -- (C2);
  \draw[inmedium] ($(B)!0.5!(D)$) circle (.3cm);
  \node at (D) {\scalebox{0.7}{$\blacktriangleright$}};
  \vertex{C}{2};
  \node at (1,0.6) { };
}

\newcommand{\CondensateTypeTwoPiPi}[0]{
  \coordinate (A) at (0,0);
  \coordinate (B) at ($(A)+(\len,0)$);
  \coordinate (C) at ($(B)+(\len,0)$);
  \coordinate (C2) at ($(C)+(\len,0)$);
  \coordinate (C3) at ($(C2)+(\len,0)$);
  \coordinate (D) at ($(C)+(0,-0.6)$);
  \draw[source] (A) -- (B);
  \draw[pion] (B) -- (C);
  \draw[pion] (C) -- (C2);
  \draw[source] (C2) -- (C3);
  \draw[inmedium] ($(C)!0.5!(D)$) circle (.3cm);
  \node at (D) {\scalebox{0.7}{$\blacktriangleright$}};
  \vertex{C}{2};
  \node at (1,0.6) { };
}

\newcommand{\CondensateTypeThreePi}[0]{
  \coordinate (A) at (0,0);
  \coordinate (B) at ($(A) + (\len,0)$);
  \coordinate (C) at ($(B) + (\len,0)$);
  \coordinate (C2) at ($(C) + (\len,0)$);
  \coordinate (E) at ($(B) + (-\len*0.7,-\len)$);
  \coordinate (F) at ($(B) + (\len*0.7,-\len)$);
  \draw[source] (A) -- (B);
  \draw[source] (C) -- (C2);
  \draw[pion] (B) -- (C);
  \draw[pion] (B) to[in=90,out=220] (E);
  \draw[pion] (F) to[in=320,out=90] (B);
  \draw[inmedium] (E) to[in=135,out=45] (F);
  \draw[inmedium] (E) to[in=225,out=-45] (F);
  \node at ($(E)!0.5!(F) + (0,0.3)$) {\scalebox{0.7}{$\blacktriangleright$}};
  \node at ($(E)!0.5!(F) + (0,-0.3)$) {\scalebox{0.7}{$\blacktriangleleft$}};
  \vertex{C}{2};
}

\newcommand{\CondensateTypeThreePiPi}[0]{
  \coordinate (A) at (0,0);
  \coordinate (Z) at ($(A) - (\len,0)$);
  \coordinate (B) at ($(A) + (\len,0)$);
  \coordinate (C) at ($(B) + (\len,0)$);
  \coordinate (C2) at ($(C) + (\len,0)$);
  \coordinate (E) at ($(B) + (-\len*0.7,-\len)$);
  \coordinate (F) at ($(B) + (\len*0.7,-\len)$);
  \draw[source] (Z) -- (A);
  \draw[pion] (A) -- (B);
  \draw[source] (C) -- (C2);
  \draw[pion] (B) -- (C);
  \draw[pion] (B) to[in=90,out=220] (E);
  \draw[pion] (F) to[in=320,out=90] (B);
  \draw[inmedium] (E) to[in=135,out=45] (F);
  \draw[inmedium] (E) to[in=225,out=-45] (F);
  \node at ($(E)!0.5!(F) + (0,0.3)$) {\scalebox{0.7}{$\blacktriangleright$}};
  \node at ($(E)!0.5!(F) + (0,-0.3)$) {\scalebox{0.7}{$\blacktriangleleft$}};
  \vertex{C}{2};
}

\newcommand{\CondensateTypeFour}[0]{
  \coordinate (A) at (0,0);
  \coordinate (B) at ($(A)+(\len,0)$);
  \coordinate (C) at ($(B)+(\len,0)$);
  \coordinate (D) at ($(C)+(\len,0)$);
  \draw[source] (A) -- (B);
  \draw[pion] (B) -- (C);
  \draw[source] (C) -- (D);
  \draw[inmedium] ($(B)!0.5!(C)$) circle (.5cm);
  \node at ($(B)!0.5!(C) + (0,0.5)$) {\scalebox{0.7}{$\blacktriangleright$}};
  \node at ($(B)!0.5!(C) - (0,0.5)$) {\scalebox{0.7}{$\blacktriangleleft$}};
}

\newcommand{\CondensateTypeFourPi}[0]{
  \coordinate (A) at (0,0);
  \coordinate (B) at ($(A)+(\len,0)$);
  \coordinate (C) at ($(B)+(\len,0)$);
  \coordinate (D) at ($(C)+(\len,0)$);
  \coordinate (E) at ($(D)+(\len,0)$);
  \draw[source] (A) -- (B);
  \draw[pion] (B) -- (C);
  \draw[pion] (C) -- (D);
  \draw[source] (D) -- (E);
  \draw[inmedium] ($(B)!0.5!(C)$) circle (.5cm);
  \node at ($(B)!0.5!(C) + (0,0.5)$) {\scalebox{0.7}{$\blacktriangleright$}};
  \node at ($(B)!0.5!(C) - (0,0.5)$) {\scalebox{0.7}{$\blacktriangleleft$}};
  \vertex{D}{2};
}

\newcommand{\CondensateTypeFourPiPi}[0]{
  \coordinate (A) at (0,0);
  \coordinate (Z) at ($(A)-(\len,0)$);
  \coordinate (B) at ($(A)+(\len,0)$);
  \coordinate (C) at ($(B)+(\len,0)$);
  \coordinate (D) at ($(C)+(\len,0)$);
  \coordinate (E) at ($(D)+(\len,0)$);
  \draw[source] (Z) -- (A);
  \draw[pion] (A) -- (B);
  \draw[pion] (B) -- (C);
  \draw[pion] (C) -- (D);
  \draw[source] (D) -- (E);
  \draw[inmedium] ($(B)!0.5!(C)$) circle (.5cm);
  \node at ($(B)!0.5!(C) + (0,0.5)$) {\scalebox{0.7}{$\blacktriangleright$}};
  \node at ($(B)!0.5!(C) - (0,0.5)$) {\scalebox{0.7}{$\blacktriangleleft$}};
  \vertex{A}{2};
  \vertex{D}{2};
}

\newcommand{\CondensateTypeFivePiPi}[0]{
  \coordinate (A) at (0,0);
  \coordinate (Z) at ($(A)-(\len,0)$);
  \coordinate (B) at ($(A)+(\len,0)$);
  \coordinate (C) at ($(B)+(\len,0)$);
  \coordinate (C2) at ($(C)+(\len,0)$);
  \coordinate (D) at ($(B)+(0,-\len)$);
  \draw[source] (Z) -- (A);
  \draw[pion] (A) -- (B);
  \draw[pion] (B) -- (C);
  \draw[source] (C) -- (C2);
  \draw[pion] (D) -- (B);
  \draw[inmedium] ($(B)!0.5!(D)$) circle (\len/2);
  \node at ($(B)!0.5!(D) + (\len/2,0)$) {\scalebox{0.7}{$\blacktriangledown$}};
  \node at ($(B)!0.5!(D) - (\len/2,0)$) {\scalebox{0.7}{$\blacktriangle$}};
  \vertex{A}{2};
  \vertex{C}{2};
}

\newcommand{\SUThreeDiagOne}[0]{
  \coordinate (A) at (0,0);
  \coordinate (B) at ($(A)+(\len,0)$);
  \coordinate (C) at ($(B)+(\len,0)$);
  \coordinate (C2) at ($(C)+(\len,0)$);
  \coordinate (D) at ($(B)+(0,-0.6)$);
  \draw[source] (A) -- (B);
  \draw[pion] (B) -- (C);
  \draw[source] (C) -- (C2);
  \draw[inmedium] ($(B)!0.5!(D)$) circle (.3cm);
  \node at (D) {\scalebox{0.7}{$\blacktriangleright$}};
  \vertex{C}{ };
  \node at (1,0.6) { };
}

\newcommand{\SUThreeDiagTwo}[0]{
  \coordinate (A) at (0,0);
  \coordinate (B) at ($(A)+(\len,0)$);
  \coordinate (C) at ($(B)+(\len,0)$);
  \coordinate (C2) at ($(C)+(\len,0)$);
  \coordinate (D) at ($(C)+(0,-0.6)$);
  \draw[source] (A) -- (B);
  \draw[pion] (B) -- (C);
  \draw[source] (C) -- (C2);
  \draw[inmedium] ($(C)!0.5!(D)$) circle (.3cm);
  \node at (D) {\scalebox{0.7}{$\blacktriangleright$}};
  \vertex{C}{ };
  \node at (1,0.6) { };
}

\newcommand{\SUThreeDiagThree}[0]{
  \coordinate (A) at (0,0);
  \coordinate (B) at ($(A)+(\len,0)$);
  \coordinate (C) at ($(B)+(\len,0)$);
  \coordinate (C2) at ($(C)+(\len,0)$);
  \coordinate (C3) at ($(C2)+(\len,0)$);
  \coordinate (D) at ($(C)+(0,-0.6)$);
  \draw[source] (A) -- (B);
  \draw[pion] (B) -- (C);
  \draw[pion] (C) -- (C2);
  \draw[source] (C2) -- (C3);
  \draw[inmedium] ($(C)!0.5!(D)$) circle (.3cm);
  \node at (D) {\scalebox{0.7}{$\blacktriangleright$}};
  \vertex{C}{ };
  \node at (1,0.6) { };
}

\newcommand{\SUThreeDiagFour}[0]{
  \coordinate (A) at (0,0);
  \coordinate (Z) at ($(A)-(\len,0)$);
  \coordinate (B) at ($(A)+(\len,0)$);
  \coordinate (C) at ($(B)+(\len,0)$);
  \coordinate (C2) at ($(C)+(\len,0)$);
  \coordinate (D) at ($(C)+(0,-0.6)$);
  \draw[source] (Z) -- (A);
  \draw[pion] (A) -- (B);
  \draw[pion] (B) -- (C);
  \draw[source] (C) -- (C2);
  \draw[inmedium] ($(C)!0.5!(D)$) circle (.3cm);
  \node at (D) {\scalebox{0.7}{$\blacktriangleright$}};
  \vertex{C}{ };
  \node at (1,0.6) { };
}

\newcommand{\SUThreeDiagFive}[0]{
  \coordinate (A) at (0,0);
  \coordinate (Z) at ($(A)-(\len,0)$);
  \coordinate (B) at ($(A)+(\len,0)$);
  \coordinate (C) at ($(B)+(\len,0)$);
  \coordinate (C2) at ($(C)+(\len,0)$);
  \coordinate (D) at ($(A)+(0,-0.6)$);
  \draw[source] (Z) -- (A);
  \draw[pion] (A) -- (B);
  \draw[pion] (B) -- (C);
  \draw[source] (C) -- (C2);
  \draw[inmedium] ($(A)!0.5!(D)$) circle (.3cm);
  \node at (D) {\scalebox{0.7}{$\blacktriangleright$}};
  \vertex{C}{ };
  \node at (1,0.6) { };
}
\begin{document}
\title{Density dependence of the quark condensate in isospin-asymmetric nuclear matter}
\author{Stephan Hübsch}
\email{huebsch@th.phys.titech.ac.jp}
\author{Daisuke Jido}
\affiliation{Department of Physics, Tokyo Institute of Technology, Meguro, Tokyo 152-8551, Japan}
\date{July 11, 2021}
\begin{abstract} 
We compute the density dependence of the quark condensate and how it changes under the influence of isospin-asymmetric nuclear matter. 
This is achieved by the calculation of the relevant Feynman diagrams using in-medium chiral perturbation theory. 
We find that the absolute value of the quark condensate decreases in nuclear matter by around 35\% at normal nuclear density, in agreement with previous model-independent calculations. 
We also present an quantitative estimation of the difference of up- and down quark in-medium condensates. 
\end{abstract}
\maketitle

\section{Introduction}
The dynamical breaking of chiral symmetry (DB$\chi$S) is an intriguing phenomenon in the low-energy spectrum of quantum chromodynamics (QCD). 
One of the order parameters of this symmetry breaking process is the quark condensate $\langle \bar qq \rangle$. 
In the symmetric phase, the value of the quark condensate is zero, and after the symmetry breaking phase transition, it acquires a finite value.
It is expected that in extreme environments of the QCD phase diagram, e.g., at high temperature and high densities, the chiral symmetry will be at least partially restored. 

One way to confirm DB$\chi$S is therefore to investigate the partial restoration of the chiral symmetry in nuclear matter. 
Recent experimental results~\cite{Suzuki2004,Friedman2004,Friedman2005}, as well as theoretical calculations based on these results~\cite{Kolomeitsev2003,Jido2008}, have revealed that the quark condensate should decrease by about $30\%$ at the saturation density. 
In fact, this decrease of the quark condensate had been predicted by a model-independent low-density relation~\cite{Drukarev1991}, according to which the sign of the experimentally available $\pi N$-$\sigma$ term determines whether the quark condensate increases or decreases in nuclear matter. 
As the $\sigma_{\pi N}$ term extracted from $\pi N$ scattering data has been found to have a positive sign~\cite{Gasser1991}, the quark condensate should decrease in nuclear matter---at least in the low-density region. 
Hitherto, it remains an open question up to which densities the low-density formalism can be applied. 

This work is an extension of a previous work~\cite{Goda2013}, where the quark condensate in symmetric nuclear matter was calculated based on the chiral Ward identity for the in-medium quark condensate proposed in Ref.~\cite{Jido2008}. 
The main distinction of the present work is, that we consider the more general case of isospin-asymmetric nuclear matter, where the densities of protons and neutrons can be varied independently. 
Previously, the so-called large $m_N$ limit, where expressions are expanded in powers of inverse nucleon masses, has been employed. 
This enables an analytical calculation of the integrals encountered, but in some cases the large $m_N$ limit leads to deviations compared to the full expression. 
In the present work we also found that more diagrams are necessary to fully describe the quark condensate. 
This includes topologically similar diagrams with different interaction vertices, as well as topologically distinct diagrams. 
These diagrams become especially important in isospin-asymmetric nuclear matter. 
In addition, we estimate the difference of the condensates for the up and down quarks in isospin-asymmetric nuclear matter in the linear density approximation.

There have been several other works that investigated the in-medium corrections to the density dependence of the quark condensate after Refs.~\cite{Oller2002,Meisner2002} introduced in-medium chiral perturbation theory. 
Reference~\cite{Kaiser2009} investigated the in-medium quark condensate in pure neutron matter, where only small deviations to the linear behavior have been reported. 
The authors furthermore suggest that a restoration of the chiral symmetry might be favored in environments with a large neutron surplus, e.g. neutron stars. 
In Ref.~\cite{Kaiser2008}, the authors used in-medium chiral perturbation theory in order to calculate the density dependence of the quark condensate in isospin-symmetric nuclear matter. 
In particular, they included one- and two-pion exchanges, as well as $\Delta$ contributions and their results are in good agreement with this work. 
Furthermore, Ref.~\cite{GomezNicola2011} investigated the temperature dependence of both $\langle \bar uu+\bar dd\rangle$ and $\langle \bar uu-\bar dd\rangle$ at zero density, as well as chiral susceptibilities, using explicit isospin breaking through $m_u\neq m_d$ and electromagnetic interactions in order to study chiral symmetry restoration. 
Let us note that the quark condensate plays an integral role in other contexts, too. 
For instance, there have been investigations on the response to a magnetic field~\cite{Hofmann2019,Hofmann2020}, or a mechanism to generate neutrino masses~\cite{Babic2021}. 
Reference~\cite{Jido2000} pointed out that the wave function renormalization of the Nambu--Goldstone boson is a significant in-medium effect when chiral symmetry is partially restored. 
Finally, we note that by using the formalism described in this work, it is also possible to calculate in-medium hadron properties, e.g., in-medium pion properties~\cite{Goda2014}. 
However, in the present scope, we will exclusively focus on the in-medium quark condensate. 

The structure of this paper is as follows. 
In \cref{sec:ward}, we explicitly show how to calculate the in-medium quark condensate $\langle\bar uu+\bar dd\rangle^*$ from the $SU(2)$ chiral Ward identity. 
In \cref{sec:ChPT,sec:Lagrangian}, we introduce in-medium chiral perturbation theory and present the Lagrangian. 
In \cref{sec:condensate}, we list the Feynman diagrams that are used to calculate the quark condensate, and also their respective integral representation. 
In \cref{sec:results} we show our numerical results for the in-medium quark condensate. 
In \cref{sec:uu-dd}, we show a quantitative estimation of $\langle\bar uu-\bar dd\rangle^*$, for which we use an $SU(3)$ chiral Lagrangian.
\Cref{sec:conclusion} is devoted to the summary and conclusion.

\section{Methods}\label{sec:methods}
We use in-medium chiral perturbation theory with an $SU(2)$ chiral Lagrangian up to second order in the chiral counting. 
The goal is to calculate the pseudoscalar two-point correlation function, which  leads us to the in-medium quark condensate as shown in \cref{sec:ward}. 
We next describe in-medium chiral perturbation theory in \cref{sec:ChPT}, and subsequently discuss our Lagrangian in \cref{sec:Lagrangian}. 
In \cref{sec:condensate}, we derive the relation between the in-medium quark condensate and the pseudoscalar two-point correlation functions. 

\subsection{In-Medium Ward Identity}\label{sec:ward}
In this section, we show how to calculate the nuclear density dependence of the in-medium quark condensate.
To do this, we follow Ref.~\cite{Goda2013} and use the chiral Ward identity.
We start with the divergence of the following time-ordered product:
\begin{align}
  \partial^\mu [\mathsf T A_\mu ^a(x) P^b(0)] &= \mathsf T [\partial^\mu A_\mu^a (x) P^b (0)] \nonumber\\ &\qquad +\delta(x^0) [A_0^a(x),P^b(0)],
\end{align}
where the axial vector current $A_\mu^a(x)$ is the Noether current of the $SU(2)$ chiral transformation and the pseudoscalar fields $P^a(x)$ are defined in terms of the quark fields as $P^a(x) = \bar q(x) \im \gamma^5 \tau^a q(x)$ with the Pauli matrices $\tau^a$ for the isospin space. 
We assume isospin symmetry for the up and down quark masses. 
We can use the partially-conserved axial current (PCAC) relation $\partial^\mu A_\mu ^a(x) = m P^a(x)$ with $m$ the quark mass, and evaluate the whole equation between the in-medium vacuum $|\Omega\rangle$, which is the ground state of the nuclear matter: 
\begin{align}
	\partial^\mu \Pi_{5\mu}^{ab}(x,0) &= m \Pi^{ab}(x,0) \nonumber\\ &\qquad + \delta(x^0) \langle\Omega|[A_0^a(x), P^b(0)]|\Omega\rangle.
\end{align}
Here we abbreviated 
\begin{align}
	\Pi_{5\mu}^{ab}(x,0) &= \langle\Omega|\mathsf T A_\mu ^a(x) P^b(0)|\Omega\rangle, \\ 
	\Pi^{ab}(x,0) &= \langle\Omega|\mathsf T P^a(x) P^b(0)|\Omega\rangle . \label{eq:Pi}
\end{align}
We perform a Fourier transformation to momentum space, 
\begin{align}
	-\im q^\mu \Pi_{5\mu}^{ab}(q) &= m \Pi^{ab}(q) \\ &\quad\;\, +\!\int\!\dd[3]\vec x\,\eu^{-\im \vec q\cdot \vec x} \langle\Omega|[A_0^a(x), P^b(0)]|\Omega\rangle, \nonumber
\end{align}
and take the soft limit $q^\mu \to 0$. 
This allows us to integrate over the last term, which gives the corresponding conserved Noether charge according to $Q^a_5 = \int \dd[3]{\vec x} A_0^a(x)$:
\begin{align}
	-\im \lim\limits_{q\to 0}q^\mu \Pi_{5\mu}^{ab}(q) &= m \Pi^{ab}(0) + \langle\Omega|[Q_5^a, P^b(0)]|\Omega\rangle.
\end{align}
In order to continue, we employ the $SU(2)$ chiral transformation behavior of the pseudoscalar current, $[Q_5^a, P^b(0)] = -\im \delta^{ab} \bar qq(0)$, to write the equation
\begin{align}\label{eq:HowToCondensate}
  m \Pi^{ab}(0) + \im \lim\limits_{q\to 0}q^\mu \Pi_{5\mu}^{ab}(q) &= \im\delta^{ab} \langle \bar qq\rangle ^*, 
\end{align}
where we have defined $\langle\Omega|\bar qq |\Omega\rangle \equiv \langle \bar qq\rangle ^* \equiv \langle \bar uu + \bar dd\rangle ^* $ in \cref{eq:HowToCondensate}. 
In the soft limit, the second term on the left-hand side vanishes, since we do not have any zero modes present. 
Conversely, in the chiral limit, the first term would vanish, because $m=0$, but the second term would survive owing to massless pion modes. 
In summary, by calculating the pseudoscalar two-point correlation function, \cref{eq:Pi}, in chiral perturbation theory, we can infer the density dependence of the in-medium quark condensate.
In summary, we use two QCD relations, PCAC and the Ward identity, in order to relate the quark condensate to pseudoscalar correlation functions, and then use chiral perturbation theory to compute those correlation functions. 

We note that it is also possible to investigate the quark condensate via the Hellmann--Feynman theorem by taking derivatives of the nucleon energy density with respect to the quark mass, see e.g. Ref.~\cite{Kaiser2008} and references therein.

\subsection{In-Medium Chiral Perturbation Theory}\label{sec:ChPT}
Chiral perturbation theory is an effective field theory, therefore its most fundamental quantity is the Lagrangian, which exhibits the symmetries of QCD. 
We follow the in-medium formalism initiated by Oller~\cite{Oller2002}, and developed by Ref.~\cite{Meisner2002}. 
Using the Lagrangian, we can construct the partition function $Z$, which contains the information of all possible interactions.
This generating functional is defined as the transition amplitude between two asymptotic \emph{in}- and \emph{out}-states, and can be written as the exponential of the functional of all connected Green functions $W$:
\begin{align}
  Z[J] = \eu^{\im W[J]} = \langle \Omega_\text{out}|\Omega_\text{in}\rangle.
\end{align}
We assume that these asymptotic states are described by Fermi seas of non-interacting protons and neutrons at times $t\to \pm\infty$. 
They can be written as $N$ excitations of the vacuum,
\begin{align}
  |\Omega\rangle = \prod\limits_{i}^N a^\dagger (\vec p_i) |0\rangle,
\end{align}
where we include all momenta up to the Fermi momentum, which depends on the nucleon density as 
\begin{align}
    k_F^{p,n} = (3\pi^2 \rho_{p,n})^{1/3}. 
\end{align}
Furthermore, $a^\dagger (\vec p_i)$ is a nucleon creation operator with momentum $\vec p_i$ and $|0\rangle$ is the zero-particle vacuum state. 
Correlations between nucleons can be implemented by an interaction Lagrangian.

In the path integral formalism, the generating functional $Z[J]$ can be expressed using the Lagrangian of the system,
\begin{eqnarray}
  Z[J] = \int \mathcal D U \mathcal D N \mathcal D N^\dagger \langle\Omega_\text{out}|N_{t\to+\infty}\rangle \nonumber\\\eu^{\im \int \dd[4]x (\mathcal L_\pi + \mathcal L_N + \mathcal L_{\pi N})} \langle N_{t\to-\infty}|\Omega_\text{in}\rangle,
\end{eqnarray}
where the full Lagrangian contains a pion term, a nucleon term and a pion--nucleon interaction term. 
The nucleon--nucleon contact interactions can be also included to describe nucleon--nucleon correlations, which can be important in $\mathcal O(\rho^2)$. 

The nucleon field appears bilinear in the Lagrangian, which means it can be integrated out~\cite{Oller2002}.  
This results in an expansion of Fermi sea insertions, 
\begin{widetext}
\begin{align} \label{eq:Z}
	Z[J] = \int \mathcal DU \exp\bigg\{ \im\int\dd[4]x \Big[ \mathcal L_{\pi} &- \int\widetilde{\dd {p}}\, \mathcal {FT} \Tr{\im\Gamma(x,y)(\slashed p + m_N)n(\vec p)} \nonumber \\
	& -\frac{\im}{2} \int \widetilde{\dd {p}}\,\widetilde{\dd {q}}\, \mathcal {FT} \Tr{\im\Gamma(x,x')(\slashed q + m_N)n(\vec q) \im\Gamma(y',y) (\slashed p + m_N) n(\vec p)} + \ldots  \Big] \bigg\} ,
\end{align}
\end{widetext}
where $\widetilde{\dd p} = \dd[3]\vec p\, (2\pi)^{-3} (2E(\vec p))^{-1}$ denotes a Lorentz invariant integration measure, $E(\vec p) = (\vec p^2 + m_N^2)^{1/2}$ is the nucleon energy for momentum $\vec p$, $\mathcal {FT}$ represents a Fourier transformation of spatial variables (except for $x$) and $\Gamma(x,y)$ is a non-local vertex defined via in-vacuum quantities, 
\begin{align}\label{eq:gamma-definition}
	\im\Gamma &= A [ \id_4 - G_0 A ]^{-1}.
\end{align}
Here, the interaction operator $A$ and the free nucleon propagator $G_0$ are given by in-vacuum chiral perturbation theory. 
The matrix $n(\vec p)$ implements the different Fermi momenta of protons and neutrons, 
\begin{align}
	n(\vec p) &= \begin{pmatrix}
		\Heavi(k_F^p - |\vec p|) & 0 \\
		0 & \Heavi(k_F^n - |\vec p|) 
	\end{pmatrix},
\end{align}
and restricts the momentum integration up to these Fermi momenta using Heaviside step functions. 
We will abbreviate these step functions as follows: $\Heavi_{\vec p}^{p,n} \equiv \Heavi(k_F^{p,n} - |\vec p|)$. 
Finally, the non-local vertex $\Gamma(x,y)$ in \cref{eq:gamma-definition} can be expanded in terms of the interaction operator $A$ and the in-vacuum nucleon propagator $G_0$ like 
\begin{align}
	\im\Gamma &= A + AG_0A + AG_0AG_0A + \ldots,
\end{align}
and the pion--nucleon interaction operator $A$ will be discussed in the next section. 

\subsection{The Chiral Lagrangian}\label{sec:Lagrangian}
We employ a chiral Lagrangian up to second order in the chiral counting. 
There are two contributions to this Lagrangian. 
We have the pion Lagrangian, 
\begin{align}\label{eq:pionLag}
	\mathcal L_\pi ^{(2)} = \frac{f^2}{4} \Tr{D_\mu U^\dagger D^\mu U + \chi^\dagger U+\chi U^\dagger},
\end{align}
which describes pion-pion interactions, as well as interactions of pions with external currents. 
The chiral field $U$ is given in terms of the pion fields $\vec\pi=\pi^a \tau^a$:
\begin{align}
	U &= \exp(\im \vec\pi \frac{y(\pi^2)}{2\sqrt{\pi^2}})\qquad (\pi^2 = \pi^a\pi^a).\label{eq:parametrization}
\end{align}
Note that the function $y(\pi^2)$ fulfills:
\begin{align}
	y-\sin(y) = \frac{4}{3} \left(\frac{\pi^2}{f^2}\right)^{3/2}.
\end{align}
This parametetrization of the chiral field enables a simple treatment  of perturbative calculations~\cite{Charap1970,Charap1971,Gerstein1971}: 
the partition function, \cref{eq:Z}, is defined in terms of the chiral field $U$, so when considering loop corrections to the pion field, one should be careful about chiral invariance of the integral measure. 
Furthermore, the Adler zero condition~\cite{Adler1965} is immediately fulfilled in this parametrization. 

The covariant derivative $D_\mu$ acts on the chiral field as 
\begin{align}
	D_\mu U &= \partial_\mu U -\im r_\mu U + \im U\ell_\mu,
\end{align}
where $r_\mu = (v_\mu+a_\mu)/2$ and $\ell_\mu = (v_\mu-a_\mu)/2$. 
In these expressions, $v_\mu=v_\mu^a \tau^a / 2$ and $a_\mu=a_\mu^a \tau^a / 2$ are the external vector- and axial-vector fields ($a=1,2,3$).

The field $\chi$ contains the external scalar ($s=s^a\tau^a$, $a=0,1,2,3$) and pseudoscalar ($p=p^a\tau^a$, $a=0,1,2,3$) currents,
\begin{equation}
	\chi = 2 B_0 (s+\im p),
\end{equation}
and $B_0$ is a low-energy constant (LEC).
The quark mass is introduced through the external scalar field by setting $s = m \tau^0$ with the quark mass $m$. 

The $\pi N$ interaction Lagrangian is given by 
\begin{align}
	\mathcal L _{\pi N} &= -\bar N A N, & A&=\sum\limits_{i=1}A^{(i)},
\end{align}
where $A^{(i)}$ is of chiral order $\mathcal O(p^i)$. 
The terms which are relevant to the present calculation read:
\begin{subequations}
\begin{align}
    A^{(1)} &= -\im \gamma^\mu \Gamma_\mu - \im g_A \gamma^\mu \gamma^5 \Delta_\mu, \\
    A^{(2)} &= -c_1 \Tr{\chi_+} + \ldots 
\end{align}
\end{subequations}
Here, we need the definitions of the vector and axial-vector currents: 
\begin{align}
    \Gamma_\mu &= \frac{1}{2} [u^\dagger, \partial_\mu u]\nonumber\\&\qquad-\frac{\im}{2} u^\dagger (v_\mu + a_\mu) u - \frac{\im}{2}u(v_\mu-a_\mu)u^\dagger, \\
    \Delta_\mu &= \frac{1}{2} u^\dagger \big[ \partial_\mu -\im(v_\mu+a_\mu) \big]u \nonumber\\&\qquad- \frac{1}{2}u\big[ \partial_\mu - \im(v_\mu-a_\mu) \big]u^\dagger,
\end{align}
where $u^2 = U$. 
The scalar and pseudoscalar sources are contained in $\chi_+$: 
\begin{align}
    \chi_+ &= u\chi^\dagger u + u^\dagger \chi u^\dagger.
\end{align}
The full definition of $A^{(1)}$ and $A^{(2)}$ are given in \cref{app:A}.

\subsection{In-Medium Quark Condensate}\label{sec:condensate}
In this section, we will calculate the in-medium quark condensate through $\Pi^{ab}(0)$ using \cref{eq:HowToCondensate}. For this purpose we will present the Feynman diagrams for the calculation of $\Pi^{ab}(0)$.
These diagrams all start and end with a wavy line, representing an external pseudoscalar isotriplet current. 

The leading contribution to the in-medium quark condensate in the density expansion reads: 
\begin{align}\label{eq:Pi1-diag}
    \Pi_1 &= \sum\limits_{i=1}^2\resizebox{\dFour}{!}{$\diagram{\CondensateTypeTwoPiPi;\vertex{B}{2};\vertex{C}{i};\vertex{C2}{2};}$}  \nonumber\\[-10pt]&\qquad +2\;
    \resizebox{\dThree}{!}{$\diagram{\CondensateTypeTwoPi;\vertex{B}{2};}$} 
\end{align}
Thick lines are in-medium nucleon propagators, dashed lines are pion propagators, wavy lines are pseudoscalar currents, and the number inside a vertex represents whether this interaction comes from $A^{(1)}$ or $A^{(2)}$. 
If no nucleon participates in an interaction, then only $\mathcal L^{(2)}$ yields a contribution. 
We have listed all necessary interaction Lagrangians in \cref{sec:Lagrangian}.
We count the last diagram in \cref{eq:Pi1-diag} twice in order to account for the fact that the in-medium nucleon loop can be attached to either vertex. 
We now show how to apply the Feynman rules which are needed to calculate these diagrams using the second diagram in \cref{eq:Pi1-diag} as an example: 
\begin{align}
    &\lim\limits_{q^\mu\to 0} (-1)^L \frac{(-1)^n}{n} (-\im)^2 \int\frac{\dd[3]\vec p}{(2\pi)^3}\frac{1}{2p_0}\\
    &\qquad \times \Tr{[-\im A_{\pi p}^{(2)}]^{ac} (\slashed p+m_N)n(\vec p)  } \frac{\im \delta^{ac}}{q^2-m_\pi^2} [\im\mathcal L_{\pi p}^{(2)}]^{cb}. \nonumber
\end{align}
Here, we have one fermionic loop, $L=1$, one Fermi-sea propagator, $n=1$, and two external currents in the correlation function, thus we get a factor of $(-\im)^2$.
Due to energy-momentum conservation, $p_0 = (\vec p^2 + m_N^2)^{1/2}$. 
The relevant terms in the Lagrangian are:
\begin{align}
  A_{\pi p}^{(2)} &= -\frac{8c_1B_0}{f}p^i\pi^i, & 
  \mathcal L_{\pi p}^{(2)} &= 2 fB_0 \pi^i p^i,
\end{align}
which implies that the trace both for isospin and Dirac indices simplifies to:
\begin{align}
    \Tr{(\slashed p+m_N)n(\vec p)  } = 4m_N^2 (\Heavi_{\vec p}^p + \Heavi_{\vec p}^n),
\end{align}
and we obtain: 
\begin{align}
	-\im\delta^{ab} \frac{16 c_1B_0^2 m_N}{\pi^2 m_\pi^2}   \int\dd p \frac{p^2}{p_0} (\Heavi_{\vec p}^p + \Heavi_{\vec p}^n).
\end{align}
As written in \cref{eq:HowToCondensate}, we multiply the pseudoscalar correlation function with the quark mass and also divide by the vacuum condensate, which is given at tree level by $\langle \bar qq \rangle_0 = -2f^2 B_0 + \ldots$, in order to obtain the contribution to the in-medium quark condensate. 
The pion mass and the quark mass are related in leading order chiral perturbation theory via $m_\pi^2 = 2mB_0+\ldots$, which fixes the value for $B_0$. 
After multiplying this diagram by $2$ and adding the other diagram in \cref{eq:Pi1-diag}, the result is: 
\begin{align}\label{eq:Pi1}
	\frac{m \Pi_{1}^{ab}(0)}{\langle \bar qq\rangle_0} &= \im\delta^{ab} \frac{4 c_1 m_N}{\pi^2 f^2} \int\dd p \frac{p^2}{p_0} [\Heavi_{\vec p}^p + \Heavi_{\vec p}^n]. 
\end{align}
This integral can be evaluated analytically: 
\begin{align}
    & \int_0^{k_F} \dd p \frac{p^2}{\sqrt{p^2+m_N^2}} =\nonumber\\&\quad  \frac{k_F}{2}\sqrt{k_F^2+m_N^2} - \frac{m_N^2}{2}\arctanh\left(\frac{k_F}{\sqrt{k_F^2+m_N^2}}\right),
\end{align}
so we can expand it in the large-$m_N$ limit as follows: 
\begin{align}
    & \frac{m_N}{\pi^2}\int\dd p \frac{p^2}{p_0} [\Heavi_{\vec p}^p + \Heavi_{\vec p}^n] = \nonumber\\&\hspace{0.5cm} \rho - \rho^{5/3} \frac{1+r^{5/3}}{10m_N^2}\left[\frac{3^5\pi^4}{(1+r)^5}\right]^{1/3} + \mathcal O\left(\frac{1}{m_N^3}\right),
\end{align}
where we abbreviate the neutron-to-proton ratio $r=\rho_n/\rho_p$. 
Since the expression in \cref{eq:Pi1} is invariant under the exchange of $k_F^p$ and $k_F^n$, it is consequently also invariant under the exchange of $\rho_p$ and $\rho_n$. 
This symmetry is also present in the other diagrams.

The next-to-leading contributions in the density expansion are given by $\Pi_2$, $\Pi_3$ and $\Pi_4$:
\begin{align}\label{eq:Pi2-diag}
    \Pi_2 &=2\;
    \resizebox{\dThree}{!}{$\diagram{\CondensateTypeThreePi;\vertex{B}{2};\vertex{E}{1};\vertex{F}{1};}$} + \resizebox{\dFour}{!}{$\diagram{\CondensateTypeThreePiPi;\vertex{A}{2};\vertex{B}{2};\vertex{E}{1};\vertex{F}{1};}$}
\end{align}
which lead to the following terms: 
\begin{widetext}
\begin{subequations}\label{eq:Pi2}
\begin{align} 
	\frac{m \Pi_{2,1}^{ab}(0)}{\langle \bar qq\rangle_0} &= \im\delta^{ab} \frac{g_A^2 m_N^2}{5 (2\pi f)^4 } \int \dd {p}\dd {k}\dd\cos\theta \frac{p^2k^2}{p_0k_0} \frac{p_0k_0-pk\cos\theta-m_N^2}{[(p-k)^2-m_\pi^2]^2}(\Heavi_{\vec p}^p\Heavi_{\vec k}^p+4\Heavi_{\vec p}^p\Heavi_{\vec k}^n+4\Heavi_{\vec p}^n\Heavi_{\vec k}^p+\Heavi_{\vec p}^n\Heavi_{\vec k}^n) \nonumber\\[4pt]
	&\qquad+ \im\delta^{a3}\delta^{b3} \frac{g_A^2 m_N^2}{10 (2\pi f)^4 m_\pi^2 } \int \dd {p}\dd {k}\dd\cos\theta \frac{p^2k^2}{p_0k_0} \frac{4m_\pi^2(p_\mu k^\mu-m_N^2)}{[(p-k)^2-m_\pi^2]^2} (\Heavi_{\vec p}^p-\Heavi_{\vec p}^n)(\Heavi_{\vec k}^p-\Heavi_{\vec k}^n),\\[12pt]
	\frac{m \Pi_{2,2}^{ab}(0)}{\langle \bar qq\rangle_0} &= \im\delta^{ab}\frac{g_A^2 m_N^2}{10 (2\pi f)^4 m_\pi^2} \int \dd {p}\dd {k}\dd \cos\theta \frac{p^2k^2}{p_0k_0}\frac{p_0k_0 - pk\cos\theta - m_N^2}{[(p-k)^2-m_\pi^2]^2} \Big[ a_1 (\Heavi_{\vec p}^p\Heavi_{\vec k}^p+\Heavi_{\vec p}^n\Heavi_{\vec k}^n) + a_2 (\Heavi_{\vec p}^p\Heavi_{\vec k}^n+\Heavi_{\vec p}^n\Heavi_{\vec k}^p) \Big]\nonumber\\[4pt]
	&\quad\; + \im\delta^{a3}\delta^{b3} \frac{g_A^2 m_N^2}{10 (2\pi f)^4 m_\pi^2} \int \dd {p}\dd {k}\dd \cos\theta \frac{p^2k^2}{p_0k_0}\frac{[2m_\pi^2-6(p-k)^2](p_\mu k^\mu - m_N^2)}{[(p-k)^2-m_\pi^2]^2}  (\Heavi_{\vec p}^p-\Heavi_{\vec p}^n)(\Heavi_{\vec k}^p-\Heavi_{\vec k}^n).
\end{align}
\end{subequations}
\end{widetext}
Here we abbreviated:
\begin{subequations}
\begin{align}
    p_\mu k^\mu &= p_0 k_0 - pk\cos\theta, \\
    a_1 &= 2(p-k)^2 + m_\pi^2,\\
	a_2 &= -2(p-k)^2 + 4m_\pi^2,
\end{align}
\end{subequations}
and can simplify $(p-k)^2=2m_N^2 - 2p_0k_0 +2pk\cos\theta$. 

The next set of diagrams cancel each other in the soft limit, where the external pion momentum goes to zero: 
\begin{align}\label{eq:Pi3-diag}
    \Pi_3 &= \resizebox{\dThree}{!}{$\diagram{\CondensateTypeFour;\vertex{B}{2};\vertex{C}{2};}$} \nonumber\\&\qquad + 2\sum\limits_{i=1}^{2} \resizebox{\dFour}{!}{$\diagram{\CondensateTypeFourPi;\vertex{B}{2};\vertex{C}{i};}$} \nonumber\\&\qquad  + \sum\limits_{i,j=1}^{2}\resizebox{\dFive}{!}{$\diagram{\CondensateTypeFourPiPi;\vertex{B}{i};\vertex{C}{j};}$}
\end{align}
The last diagram we consider for the next-to-leading order in the density expansion is: 
\begin{align}\label{eq:Pi4-diag}
    \Pi_4 = \resizebox{\dFour}{!}{$\diagram{\CondensateTypeFivePiPi;\vertex{B}{1};\vertex{D}{1};}$}
\end{align}
which results in:
\begin{align}\label{eq:Pi4}
    \frac{m \Pi_{4}^{ab}(0)}{\langle \bar qq\rangle_0} &= \im \frac{g_A^2 m_N^2}{10(2\pi f)^4 m_\pi^2} \int \dd {p} \dd {k} \dd\cos\theta \frac{p^2k^2}{p_0k_0}\nonumber\\&\qquad \times \frac{p_\mu k^\mu-m_N^2}{(p-k)^2-m_\pi^2 } \nonumber\\&\qquad \times (\Heavi_{\vec p}^p-\Heavi_{\vec p}^n)(\Heavi_{\vec k}^p-\Heavi_{\vec k}^n)\nonumber\\&\qquad \times (-2\delta^{ab} + 6 \delta^{a3}\delta^{b3} ).
\end{align}

According to \cref{eq:HowToCondensate}, the pseudoscalar correlation function $\Pi^{ab}$ is only proportional to $\delta^{ab}$. 
Still, we also found contributions proportional to $\delta^{a3}\delta^{b3}$ in \cref{eq:Pi2,eq:Pi4}, but after adding all diagrams, the contributions proportional to $\delta^{a3}\delta^{b3}$ exactly cancel each other, in accordance with chiral symmetry. 
Otherwise, this would violate the chiral $SU(2)$ symmetry because the third isospin component of the pseudoscalar current $P^a$ would behave in a different way compared to the other two components: $\Pi^{11}=\Pi^{22}\neq \Pi^{33}$. 

\section{Results}\label{sec:results}
In this section, we show our numerical results. 
In our calculations, we use the following numerical values for the low-energy constant $c_1 = -0.59$ GeV${}^{-1}$, the pion decay constant $f_\pi = 92.4$ MeV, and the axial coupling $g_A = 1.26$~\cite{Goda2013}. 
We also use isospin-averaged values for the pion mass, $m_\pi = 138$ MeV, and the nucleon mass, $m_N=939$ MeV. 

\begin{figure}
    \centering
    \includegraphics[width=0.48\textwidth]{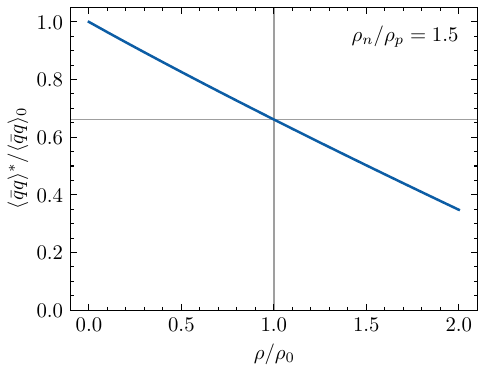}
    \caption{Density dependence of the in-medium quark condensate, normalized to the vacuum condensate. The ratio of neutrons to protons is given by $\rho_n/\rho_p$. At normal nuclear density, $\rho=\rho_0$, the quark condensate is reduced by about $35\%$ compared to its vacuum value. }
    \label{fig:condensate}
\end{figure}

For the in-medium quark condensate, the integrals in \cref{eq:Pi1,eq:Pi2,eq:Pi4} are solved numerically and the result yields the density dependence of the quark condensate, which is presented in \cref{fig:condensate}. 
Our computed value of the quark condensate at normal nuclear density, which shows a reduction of $34.3\%$ shows good agreement with Ref.~\cite{Jido2008}, where the in-medium behavior of the quark condensate was shown to be: $\flatfrac{\langle\bar qq\rangle^*}{\langle\bar qq\rangle_0}\approx (\flatfrac{b_1}{b_1^*})^{1/2} [1-\tfrac{1}{2}\beta\rho] \approx 1-0.37\, \rho/\rho_0$.  
Here, the parameters take on the following values $b_1/b_1^*|_{\rho=\rho_0}\approx 0.79\pm 0.05$ and $\beta\approx (2.17\pm 0.04)\text{ fm}^3$, which were obtained from experimental data from deeply bound pionic atoms and isospin-singlet $\pi N$-scattering amplitudes, respectively. 
Our results are also in good agreement with results obtained in Ref.~\cite{Kaiser2008} which are summarized by Ref.~\cite{Gubler2019}.
They are also consistent with Ref.~\cite{Cohen1992}. 

\begin{figure*}
    \centering
    \includegraphics[width=0.99\textwidth]{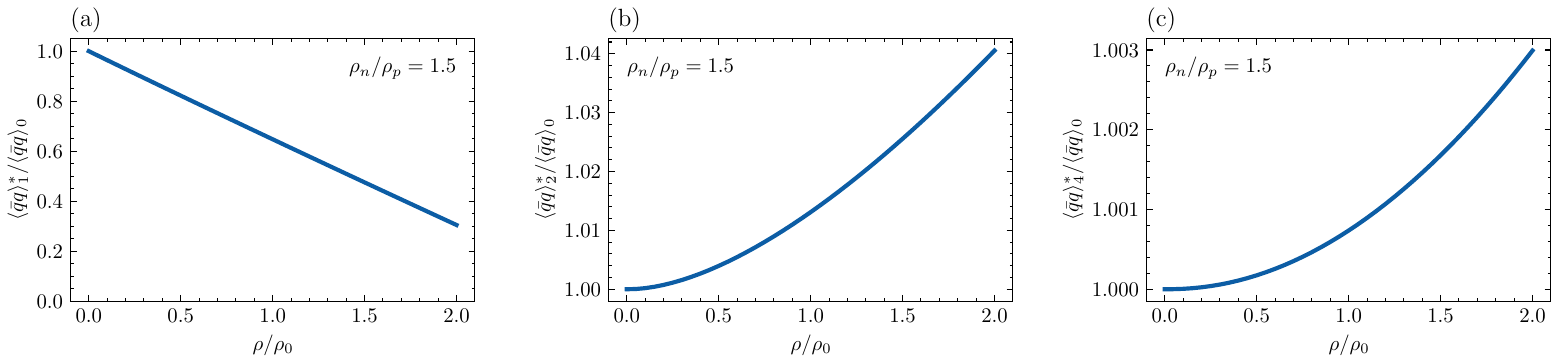}
    \caption{Density dependence of the contributions to the in-medium quark condensate, normalized to the vacuum condensate. The ratio of neutrons to protons is given by $\rho_n/\rho_p$. \cref{fig:condensate} shows the sum of these three contributions. (a) The leading order contributions corresponding to \cref{eq:Pi1}. (b) The next-to-leading order contributions corresponding to \cref{eq:Pi2}. (c) The next-to-leading order contributions corresponding to \cref{eq:Pi4}. }
    \label{fig:condensate-contributions}
\end{figure*}

The density dependence of the in-medium quark condensate shows a linear behavior, since the next-to-leading order contributions in the density expansion, which lead to $\mathcal O(\rho^{4/3})$ and $\mathcal O(\rho^{5/3})$ behavior, are much smaller. 
The contributions of the individual diagrams are shown in \cref{fig:condensate-contributions}.  
The right two figures show that the next-to-leading order diagrams have rather tiny contributions to the quark condensate. 
This is because the Fermi momentum is small compared to the nucleon mass in low densities and the Fermi motion expansion could be convergent. 
In this sense, the nucleon-nucleon correlation, which plays a role from $\mathcal O(\rho^2)$, should be important and bring a new scale parameter. 

This work is built on the foundation of Ref.~\cite{Goda2013}, where the in-medium quark condensate was calculated for isospin-symmetric nuclear matter. 
In this work, we found several new diagrams which are necessary contributions to the in-medium quark condensate. 
In particular, the first diagram in \cref{eq:Pi1-diag} with $i=1$, the second diagram in \cref{eq:Pi3-diag} with $i=1$ as well as the third diagram in \cref{eq:Pi3-diag} with $i\neq j$. 
Furthermore, diagrams like the one in \cref{eq:Pi4-diag} were not considered at all in Ref.~\cite{Goda2013}. 
Comparing to Ref.~\cite{Kaiser2008} and related works, we include interactions from $A^{(2)}$, but omit $2\pi$-exchange and $\Delta$-excitation processes, as they contribute beyond $\mathcal O(\rho^{5/3})$. In order to include such processes, Ref.~\cite{Kaiser2008} uses a nucleon-nucleon potential obtained from lattice QCD data.

\begin{figure}
    \centering
    \includegraphics[width=0.48\textwidth]{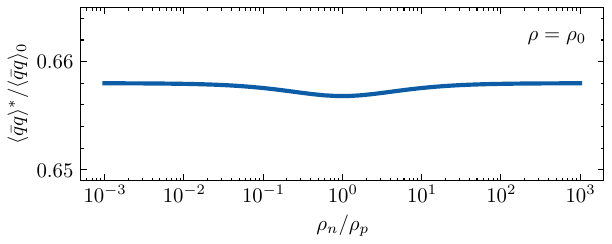}
    \caption{Density dependence of the in-medium quark condensate, normalized to the vacuum condensate. The ratio of neutrons to protons is given by $\rho_n/\rho_p$. The in-medium quark condensate at a fixed density is symmetric around the point $\rho_n/\rho_p=1$ under the exchange $\rho_n\leftrightarrow\rho_p$. The far left of this plot corresponds to proton matter, whereas the far right of this plot corresponds to neutron matter.}
    \label{fig:condensate-different-ratios}
\end{figure}

Despite being a small contribution, the isospin-asymmetry of the surrounding nuclear matter plays a role in the in-medium quark condensate. 
As shown in \cref{fig:condensate-different-ratios}, for normal nuclear density $\rho = \rho_0$, different values for the neutron-to-proton ratio have a small effect on the in-medium quark condensate. 
In particular, when plotted logarithmically, one can clearly see that the plot is symmetric around the point $\rho_n/\rho_p=1$, which symbolizes isospin symmetric nuclear matter. 
This means, the in-medium quark condensate behaves the same for a certain ratio $\rho_n/\rho_p$ and for the inverse ratio $(\rho_n/\rho_p)^{-1}$ because of isospin symmetry. 
This can also be seen on the far left and far right ends of the plot, which correspond to proton matter and neutron matter, respectively. 
Furthermore, there is no significant effect if we treat protons and neutrons with different masses $m_p\neq m_n$. 

For interactions between nucleons the $\Delta$ baryons might be an important ingredient, as suggested by Ref.~\cite{Kaiser2008}, however in this work we do not consider dynamical $\Delta$ interactions. 
Since some $\Delta$ contributions are already implicitly included in the low-energy constants, our calculations are expected to be valid as long as dynamical $\Delta$ interactions are not required. 
Otherwise, one would have to include the $\Delta$ fields in the Lagrangian and investigate their in-medium effects.

\section{Estimating the Difference of the Quark Condensates}\label{sec:uu-dd}
Finally, we would like to present an estimation of the isospin splitting of the quark condensate in nuclear matter, i.e., the quantity $\langle \bar uu-\bar dd\rangle^*$ in leading order of the density expansion. 
This splitting arises due to explicit isospin breaking, which we will consider via $m_u\neq m_d$ in the Lagrangian and $\rho_p\neq \rho_n$ in the nuclear matter. 
There is also isospin breaking via different hadron masses, like $m_p\neq m_n$, however since these effects yield minor corrections, $\flatfrac{(m_n-m_p)}{(m_n+m_p)}\approx 10^{-4}$, we neglect them, and instead use isospin-averaged masses. 
We also omit $SU(3)$ breaking in the meson decay constant. 

\begin{figure*}
    \centering
    \includegraphics[width=0.99\textwidth]{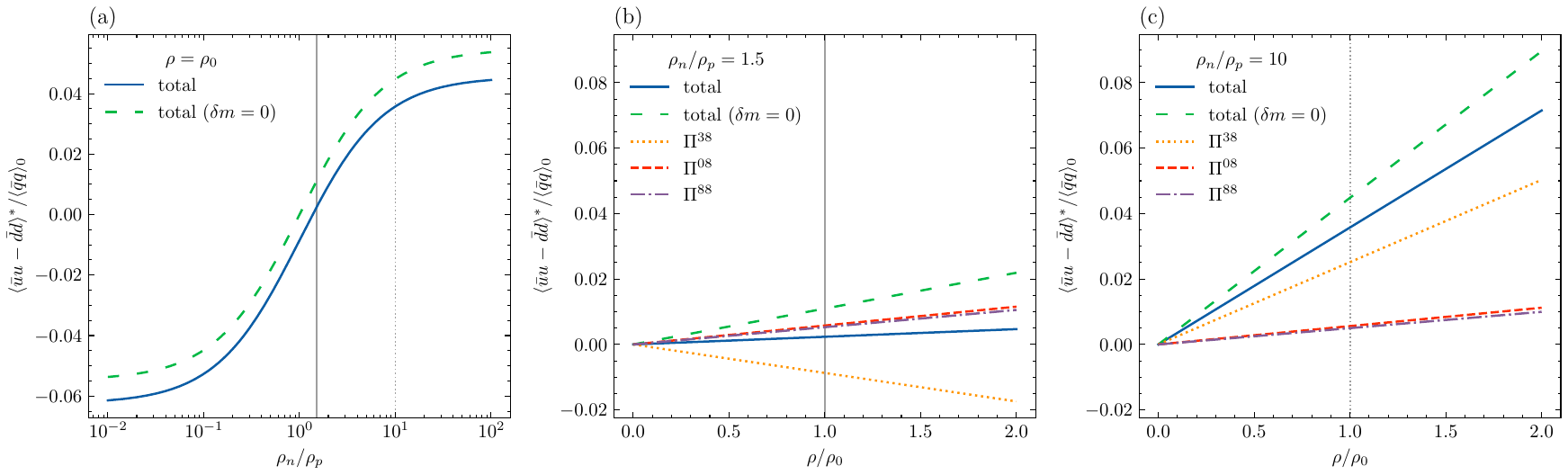}
    \caption{The density dependence of $\langle\bar uu-\bar dd\rangle^*$ is one order of magnitude smaller than the density dependence of $\langle\bar uu+\bar dd\rangle^*$. The vertical grey solid and dotted lines correspond to the same densities and nucleon ratios across the three plots. The dotted green lines correspond to the effects of only the isospin-asymmetric nuclear matter. (a) The dependence of $\langle\bar uu-\bar dd\rangle^*$ on the neutron-to-proton ratio. The isospin breaking due to $\delta m\neq 0$ leads to a reduction of around $1\%$ at normal nuclear density, across all nucleon ratios. (b) The dependence of $\langle\bar uu-\bar dd\rangle^*$ on the nucleon density. The three contributions to $\langle\bar uu-\bar dd\rangle^*$ as well as their sum are shown. For a nucleon ratio of $\rho_n/\rho_p=1.5$, the up and down quark condensates are almost the same. (c) Same as before, here with a ratio $\rho_n/\rho_p=10$. }
    \label{fig:uu-dd}
\end{figure*}

The present calculation is an estimate for two reasons. 
First, the values of the LECs are not fixed by scattering data, and second, there are no dynamic meson loops included, i.e., this is a linear-density approximation. 
In order to compute $\langle \bar uu-\bar dd\rangle^*$ in the same way as we did for $\langle \bar uu+\bar dd\rangle^*$, we need to extend our formalism to $SU(3)$, since the $SU(2)$ chiral transformation of a pseudoscalar current can only yield the sum $\bar uu+\bar dd$. 
We note that this is not the only way to compute the density dependence of the condensate difference. 
By including explicit isospin breaking via $m_u\neq m_d$, one can also use the scalar correlator $\langle\Omega|S^aS^b|\Omega\rangle$ (see e.g., Ref.~\cite{Gomez2018}).
Another way is to consider the Ward identity for the $\Pi^{30}_{5\mu}$ correlator, which would require an extension to $U(2)$, in order to include a two-flavor eta meson. 
However, for a proper definition of the eta meson, it is natural to go to the $SU(3)$ formalism.

The $SU(3)$ current algebra between an axial-vector charge and a pseudoscalar current is given by: 
\begin{align}
	[Q^a_5, P^b(0)] = -\im \sqrt{\frac{2}{3}} \delta^{ab}S^0(0) -\im d^{abc} S^c(0).
\end{align}
Here, $d^{abc}$ are the totally symmetric $SU(3)$ structure constants, given by $4d^{abc}=\Tr{\lambda^a \{\lambda^b,\lambda^c\}}$, where $\lambda^a$ ($a=1,\ldots,8$) are the Gell-Mann matrices. 
For specific values of $a$ and $b$, this yields: 
\begin{align}
    [Q^3_5, P^8(0)] &= -\frac{\im}{\sqrt 3} (\bar uu - \bar dd).
\end{align}
In the case of explicit isospin breaking in the Lagrangian due to $m_u\neq m_d$, the PCAC relation changes as follows: 
\begin{align}
	\partial^\mu A_\mu ^3(x) &= m P^3 + \delta m \sqrt{\frac{2}{3}} P^0 + \frac{\delta m}{\sqrt 3} P^8.
\end{align}
where $m=(m_u+m_d)/2$ and $\delta m= (m_u-m_d)/2$. 
This leads to two additional correlation functions necessary in order to compute the density dependence of the quark condensate compared to the $\delta m=0$ case: 
\begin{align}\label{eq:uu-dd-explicit-breaking}
	\langle \bar uu-\bar dd\rangle^* &= -\im\sqrt{3} \,m\,\Pi^{38}(0) \nonumber\\
	&\qquad -\im\sqrt{2}\,\delta m\,\Pi^{08}(0) -\im\,\delta m\, \Pi^{88}(0).
\end{align}
To this end, we deviced a similar derivation as shown in \cref{sec:condensate}. 

We use the following Lagrangian:
\begin{align} 
	\mathcal L &= \frac{f^2}{4} \Tr{D_\mu U^\dagger D^\mu U + \chi^\dagger U+\chi U^\dagger} \nonumber \\
    &\quad + b_D \Tr{\bar B\{\chi_+, B\}} + b_F \Tr{\bar B[\chi_+, B]} \nonumber \\
    &\quad + b_0 \Tr{\bar BB}\Tr{\chi_+} + \ldots, \label{eq:SU3Lag}
\end{align}
where we listed only the terms relevant to this work. 
One can identify the following relation between $SU(2)$ and $SU(3)$ low energy constants: $2b_0 + b_D + b_F = 2c_1$. 
Furthermore, there is also a $\pi^0\eta_8$ interaction Lagrangian, 
\begin{align}
	\mathcal L_{\pi^0 \eta} &= -\frac{2 B_0\, \delta m}{\sqrt{3}} \pi^0 \eta_8,
\end{align}
which allows for $\pi^0$-$\eta_8$ mixing. 
We regard this mixing as a perturbation and include relevant vertices in the Feynman diagrams. 
We will in the following consider all diagrams with additional pions coming from this $\pi^0$-$\eta_8$ mixing, since the large mass of the $\eta_8$ leads to a suppression of new $\eta_8$ propagators in the soft limit. 

For the chiral field, we use the Coleman--Callan--Wess--Zumino (CCWZ) parametrization: 
\begin{align}
    U=\exp(\im \sqrt{2}\Phi /f),
\end{align}
where the meson octet is given as $\Phi = \Phi^a \lambda^a /\sqrt 2$:  
\begin{align}
    \Phi = \begin{pmatrix}
        \frac{\pi^0}{\sqrt 2} + \frac{\eta_8}{\sqrt 6} & \pi^+ & K^+ \\[6pt]
		\pi^- & -\frac{\pi^0}{\sqrt 2} + \frac{\eta_8}{\sqrt 6} & K^0 \\[6pt]
		K^- & \bar K^0 & -\frac{2}{\sqrt 6}\eta_8
    \end{pmatrix},
\end{align}
and the baryon octet as $B=B^a\lambda^a/\sqrt 2$:
\begin{align}
    B &= \begin{pmatrix}
		\frac{\Sigma^0}{\sqrt 2} + \frac{\Lambda}{\sqrt 6} & \Sigma^+ & p \\[6pt]
		\Sigma^- & -\frac{\Sigma^0}{\sqrt 2} + \frac{\Lambda}{\sqrt 6} & n \\[6pt]
		\Xi^- & \Xi^0 & -\frac{2}{\sqrt 6}\Lambda
	\end{pmatrix}.
\end{align}
The $SU(3)$ equivalent of the chiral field parametrization in \cref{eq:parametrization} is not known, but since this linear density approximation does not include meson loops, the simpler CCWZ parametrization suffices.

The interaction terms in the Lagrangian that are necessary for the following calculations are listed in \cref{app:lagrangian}. 
The diagrams and their results are: 
\begin{align}
    \Pi^{38}(q) &= \resizebox{\dFour}{!}{$\diagram{
        \SUThreeDiagOne;
        \vertex{B}{ };
        \node[left] at (A) {$P^8$};
        \node[right] at (C2) {$P^3$};
        \node[above] at ($(B)!0.5!(C)$) {$\pi^0$};
        }$} \nonumber \\[-8pt]
    &\qquad + \resizebox{\dFour}{!}{$\diagram{
        \SUThreeDiagTwo;
        \vertex{B}{ };
        \node[left] at (A) {$P^8$};
        \node[right] at (C2) {$P^3$};
        \node[above] at ($(B)!0.5!(C)$) {$\eta_8$};
        }$}\nonumber \\[-8pt]
    &\qquad + \resizebox{\dFive}{!}{$\diagram{
    	\SUThreeDiagFive;
		\vertex{A}{ };
		\vertex{B}{ };
		\node[left] at (Z) {$P^8$};
        \node[right] at (C2) {$P^3$};
        \node[above] at ($(A)!0.5!(B)$) {$\eta_8$};
        \node[above] at ($(B)!0.5!(C)$) {$\pi^0$};
		}$}\nonumber \\[-8pt]
    &\qquad + \resizebox{\dFive}{!}{$\diagram{
        \SUThreeDiagThree;
        \vertex{B}{ };
        \vertex{C2}{ };
        \node[left] at (A) {$P^8$};
        \node[right] at (C3) {$P^3$};
        \node[above] at ($(B)!0.5!(C)$) {$\eta_8$};
        \node[above] at ($(C)!0.5!(C2)$) {$\pi^0$};
        }$}\nonumber \\[-8pt]
    &\qquad + \resizebox{\dFive}{!}{$\diagram{
    	\SUThreeDiagFour;
		\vertex{A}{ };
		\vertex{B}{ };
		\node[left] at (Z) {$P^8$};
        \node[right] at (C2) {$P^3$};
        \node[above] at ($(A)!0.5!(B)$) {$\eta_8$};
        \node[above] at ($(B)!0.5!(C)$) {$\pi^0$};
		}$}
\end{align}
The isospin breaking in the first and second diagrams is due to $\rho_p\neq \rho_n$, in the third and fifth diagrams it is due to $\delta m\neq 0$, and the fourth diagram contains terms from both. 
Furthermore, the second diagram and the term due to $\rho_p\neq \rho_n$ in the fourth diagram exactly cancel each other, because the pion mass from the propagator and from the LEC $B_0$ cancelled out. 
The correlation function $\Pi^{08}$ leads to two diagrams: 
\begin{align}
    \Pi^{08}(q) &= \resizebox{\dFour}{!}{$\diagram{
        \SUThreeDiagTwo;
        \vertex{B}{ };
        \node[left] at (A) {$P^8$};
        \node[right] at (C2) {$P^0$};
        \node[above] at ($(B)!0.5!(C)$) {$\eta_8$};
        }$}\nonumber \\[-8pt]
    &\qquad + \resizebox{\dFive}{!}{$\diagram{
    	\SUThreeDiagFour;
		\vertex{A}{ };
		\vertex{B}{ };
		\node[left] at (Z) {$P^8$};
        \node[right] at (C2) {$P^0$};
        \node[above] at ($(A)!0.5!(B)$) {$\eta_8$};
        \node[above] at ($(B)!0.5!(C)$) {$\pi^0$};
		}$}
\end{align}
These diagrams must be isospin-even. 
That is because we multiply $\Pi^{08}$ with $\delta m$, as stated in \cref{eq:uu-dd-explicit-breaking}. 
Here, the first diagram contains no isospin breaking. 
The second diagram breaks isospin twice, once via $\delta m\neq 0$ and once via $\rho_p\neq\rho_n$. 
Hence, this diagram yields a contribution of order $\mathcal O(\delta m^2)$ to the quark condensate. 
The pseudoscalar correlation function $\Pi^{88}$ can be calculated using these diagrams: 
\begin{align}
    \Pi^{88}(q) &= 2\resizebox{\dFour}{!}{$\diagram{
        \SUThreeDiagTwo;
        \vertex{B}{ };
        \node[left] at (A) {$P^8$};
        \node[right] at (C2) {$P^8$};
        \node[above] at ($(B)!0.5!(C)$) {$\eta_8$};
        }$}\nonumber \\[-8pt]
    &\qquad + 2\resizebox{\dFive}{!}{$\diagram{
    	\SUThreeDiagFour;
		\vertex{A}{ };
		\vertex{B}{ };
		\node[left] at (Z) {$P^8$};
        \node[right] at (C2) {$P^8$};
        \node[above] at ($(A)!0.5!(B)$) {$\eta_8$};
        \node[above] at ($(B)!0.5!(C)$) {$\pi^0$};
		}$}\nonumber \\[-8pt]
    &\qquad + \resizebox{\dFive}{!}{$\diagram{
        \SUThreeDiagThree;
        \vertex{B}{ };
        \vertex{C2}{ };
        \node[left] at (A) {$P^8$};
        \node[right] at (C3) {$P^8$};
        \node[above] at ($(B)!0.5!(C)$) {$\eta_8$};
        \node[above] at ($(C)!0.5!(C2)$) {$\eta_8$};
        }$}
\end{align}
These diagrams must again be isospin-even, since we multiply $\Pi^{88}$ with $\delta m$, as stated in \cref{eq:uu-dd-explicit-breaking}. 
The first diagram contains no isospin breaking, whereas the second diagram features isospin breaking due to $\delta m\neq 0$ and $\rho_p\neq\rho_n$. 
The third diagram contains one term without any isospin breaking and one with double isospin breaking due to $\delta m\neq 0$ and $\rho_p\neq\rho_n$. 

In the soft limit and omitting terms beyond $\mathcal O(\rho)$ and $\mathcal O(\delta m)$, which means omitting terms with triple isospin breaking, we get: 
\begin{subequations}
\begin{align}
	\sqrt{3}\frac{m \Pi^{38}}{\im \langle\bar qq\rangle_0} &= \frac{2\rho}{f^2} (b_D+b_F) \frac{1-r}{1+r}  \nonumber\\
	&\quad -\frac{2\rho}{f^2} (18 b_0+11 b_D+3 b_F)\frac{\delta m}{m+2m_s}, \\
	\sqrt{2}\frac{\delta m \Pi^{08}}{\im\langle\bar qq\rangle_0} &= -\frac{4 \rho (b_D-3 b_F)}{f^2} \frac{\delta m}{m+2m_s}, \\
	\frac{\delta m \Pi^{88}}{\im\langle\bar qq\rangle_0} &= \frac{\rho}{f^2} \left[ 2 (9 b_0+7 b_D-3 b_F)+\frac{m_\pi^2}{m_\eta^2}(b_D - 3b_F) \right] \nonumber\\
	&\quad \times\frac{\delta m}{m+2m_s} .
\end{align}
\end{subequations}

In agreement with the Vafa--Witten theorem~\cite{Vafa1984}, our results only depend on the difference of proton and neutron densities in the absence of explicit isospin breaking ($\delta m=0$), since isospin symmetry is not spontaneously broken in QCD.

For the LECs in the $SU(3)$ Lagrangian, we chose $b_D=0.225$ GeV$^{-1}$, $b_F = {-}0.404$ GeV$^{-1}$ and $b_0={-}0.609$ GeV$^{-1}$, which were determined in Ref.~\cite{Ren2012,Geng2013}. 
There, the values are listed following a calculation of octet baryon masses using an $SU(3)$ chiral perturbation theory with the extended-on-mass-shell renormalization scheme and fitting the low energy constants to lattice data.
Using lattice data, it is possible to determine the parameters $m_0$ (the octet baryon masses in the chiral limit) and $b_0$ separately. 
Similar values have been reported in Refs.~\cite{Aoki2017,Holmberg2018,Kubis2001}. 
We further used a quark mass ratio of $\delta m / m \approx {-}1/3$ (equivalent to $m_u/m_d\approx 0.46$) and $m_s/m \approx 27.2$ and the following leading order relations of meson masses with relation to $B_0$: 
\begin{align}
	m_\pi^2 &= 2B_0 m, & m_\eta^2 &= \frac{2}{3} B_0 (m+2m_s),
\end{align}
with $m_\pi = 138$ MeV and $m_\eta = 548$ MeV. 

The results of $\Pi^{38}$, $\Pi^{08}$ and $\Pi^{88}$ are shown in \cref{fig:uu-dd}. 
The density dependence of $\langle\bar uu-\bar dd\rangle^*$ is one order of magnitude smaller than the density dependence of $\langle\bar uu+\bar dd\rangle^*$. 
This is because the coefficient in $\langle\bar uu-\bar dd\rangle^*$ is smaller by a factor of $10$ than the coefficient in $\langle\bar uu+\bar dd\rangle^*$, as
\begin{subequations}
\begin{align}
	b_D + b_F &\approx -0.18\text{ GeV}^{-1}, \\
	2b_0 + b_D + b_F &\approx -1.4\phantom{0} \text{ GeV}^{-1}.
\end{align}
\end{subequations}
We illustrate the effects of explicit isospin breaking by also plotting the result for $\delta m=0$. 
\Cref{fig:uu-dd}~(a) shows the dependence of $\langle\bar uu-\bar dd\rangle^*$ on the neutron-to-proton ratio. 
The isospin breaking due to $m_u\neq m_d$ leads to a reduction of around $1\%$ at normal nuclear density, almost independent of nucleon ratios. 
\Cref{fig:uu-dd}~(b) shows the dependence of $\langle\bar uu-\bar dd\rangle^*$ on the nucleon density, in particular the three contributions to $\langle\bar uu-\bar dd\rangle^*$ as well as their sum are shown. 
For a nucleon ratio of $\rho_n/\rho_p=1.5$, the up and down quark condensates are almost the same. 
\Cref{fig:uu-dd}~(c) also shows the density dependence and was calculated using a nucleon ratio $\rho_n/\rho_p=10$. 

Although the explicit isospin breaking due to non-equal quark masses in the Lagrangian provides a numerically smaller effect than the isospin breaking from the surrounding nuclear matter at e.g. $\rho_n/\rho_p=1.5$, 
\begin{align}
	\frac{1}{150}\approx \left|\frac{\delta m}{m+2m_s}\right| \ll \left|\frac{1-r}{1+r}\right|=\frac{1}{5},
\end{align}
this is compensated by the large number of diagrams due to $\delta m\neq 0$, which leads to a large contribution via the LECs: 
\begin{subequations}
\begin{align}
	{-}(b_D+b_F) \frac{1-r}{1+r} &\approx 0.04\text{ GeV}^{-1}, \label{eq:compare1}\\
	(18 b_0+11 b_D+3 b_F)\frac{\delta m}{m+2m_s} &\approx 0.06\text{ GeV}^{-1}. \label{eq:compare2}
\end{align}
\end{subequations}
Here we note that the coefficients of the LECs in \cref{eq:compare2} are one order of magnitude larger than the ones in \cref{eq:compare1}. 
This is significant, especially since $b_0$ is larger than the other LECs $b_D$ and $b_F$, which gets further enhanced by the factor of $18$. 
Hence, the effects of isospin breaking in the Lagrangian and in the nuclear matter are of similar size in our calculations. 
Still, there exists a possible ambiguity in the determination of $b_0$, which is often absorbed in the chiral limit octet baryon masses $m_0$. 
It is possible to extract a value for $b_0$ from scattering experiments, although such experimental data is scarce. 
Such an issue is absent in the LECs $b_D$ and $b_F$, which are determined via Gell-Mann--Okubo mass relations. 

For a nucleon ratio of $1.5$, i.e. the one most accessible to experiments via heavy nuclei, the up and down quark condensates behave almost the same with increasing density, see \cref{fig:uu-dd} (a). 
The effect of explicit isospin breaking in the Lagrangian leads to an almost constant reduction of $1\%$ along all nucleon ratios, but the splitting of up and down quark condensates increases due to the isospin breaking of the surrounding nuclear matter. 

It would be interesting to discuss the experimental determination of the coefficients $b_D + b_F$ etc., appearing in our results. 
As is well-known, the parameter $c_1$ is determined by the $\pi N$-$\sigma$ term, which is obtained by taking the soft limit in $\pi N$ scattering, $\lim_{p\to 0} T_{\pi N}(p) = - \sigma_{\pi N}/f^2$. 
In a similar way, the $SU(3)$ coefficients could be extracted from the $\eta N \to \pi^0 N $ scattering amplitudes in the soft limit. 
Nevertheless, it would be difficult, because the $\eta$ meson mass is so large that a theoretical extrapolation to the soft limit would have a large uncertainty. 
In addition, in $\eta N \to \pi^0 N$ scattering, the contribution of the $N(1535)$ nucleon resonance is known to dominate the amplitude around the threshold. 
The resonance contribution should therefore be counted in the extrapolation. 

\section{Conclusion}\label{sec:conclusion}
We have calculated the density dependence of the $\langle \bar uu + \bar dd \rangle$ quark condensate in isospin-asymmetric nuclear matter using an SU(2) in-medium chiral perturbation theory up to second order in the chiral counting. 
We have found that the reduction of the magnitude of the quark condensate in an isospin asymmetric nuclear matter with $\rho_n/\rho_p=1.5$ agrees with the phenomenological result obtained using experimental observations and also previous theoretical estimations. 
We have also found that the effect of the isospin-asymmetric nuclear matter is weak, but still non-zero. 
This is because contributions beyond the linear density are rather small due to smaller Fermi motion in low densities. 
Nucleon-nucleon correlations, which play a role at ${\cal O}(\rho^2)$, could contribute to the deviation of the $\langle \bar uu + \bar dd \rangle$ quark condensate in the asymmetric nuclear matter from that in the symmetric nuclear matter. 

We have also estimated the difference of the up and down quark condensates, $\langle \bar uu - \bar dd \rangle^*$, in leading order of the in-medium chiral perturbation theory by using an SU(3) chiral Lagrangian. 
The isospin asymmetric nuclear matter as well as $m_u\neq m_d$ in the Lagrangian provided the splitting of the the up and down quark condensates. 
We have found, nevertheless, that the magnitude of the spitting is not so large in comparison with the in-medium reduction of the quark condensate $\langle \bar qq \rangle^*$. 

Future works might include nucleon-nucleon interaction terms in the Lagrangian. 
This would lead to new effects beyond the linear density and a new scale for the density expansion. 
It should be certainly interesting to perform dynamical calculations in the SU(3) in-medium chiral perturbation theory beyond the linear density approximation. 

\section*{Acknowledgements}
We would like to thank Prof.~Hosaka for his suggestion to calculate the in-medium density dependence of $\langle \bar uu-\bar dd\rangle$.
The work of D.\,J.~was partly supported by Grants-in-Aid for Scientific Research from JSPS (17K05449, 21K03530).

\appendix

\section{Pion-Nucleon Interaction Operator}\label{app:A}
The $\pi N$ interaction Lagrangian is given by $A=A^{(1)} + A^{(2)}+\ldots $. 
where $A^{(i)}$ is of chiral order $\mathcal O(p^i)$. 
The first two terms, $A^{(1)}$ and $A^{(2)}$ read:
\begin{align}
	A^{(1)} &= -\im \gamma^\mu \Gamma_\mu -\im g_A \gamma^\mu \gamma^5 \Delta_\mu, \label{eq:A1}\\
	A^{(2)} &= -c_1 \Tr{\chi_+} + \frac{c_2}{2m_N^2}\Tr{u_\mu u_\nu} D^\mu D^\nu \nonumber \\
	&\qquad- \frac{c_3}{2}\Tr{u_\mu u^\mu} + \frac{c_4}{2} \gamma^\mu \gamma^\nu [u_\mu, u_\nu]- c_5 \hat\chi_+  \nonumber \\
	&\qquad- \frac{\im c_6}{8m_N} \gamma^\mu \gamma^\nu F_{\mu\nu}^+ - \frac{\im c_7}{8m_N} \gamma^\mu\gamma^\nu \Tr{F_{\mu\nu}^+} . \label{eq:A2}
\end{align}
Here, we need the definitions of the following terms.
The vector current reads:
\begin{align}
    \Gamma_\mu &= \frac{1}{2} [u^\dagger, \partial_\mu u]\nonumber\\&\qquad-\frac{\im}{2} u^\dagger (v_\mu + a_\mu) u - \frac{\im}{2}u(v_\mu-a_\mu)u^\dagger, 
\end{align}
and the axial-vector current is given by: 
\begin{align}
    \Delta_\mu &= \frac{1}{2} u^\dagger \big[ \partial_\mu -\im(v_\mu+a_\mu) \big]u \nonumber\\&\qquad- \frac{1}{2}u\big[ \partial_\mu - \im(v_\mu-a_\mu) \big]u^\dagger,
\end{align}
where $u^2 = U$. 
Also, $u_\mu$ is proportional to the axial-vector current, $u_\mu = 2\im \Delta_\mu$. 
The next terms contain scalar and pseudoscalar sources, 
\begin{align}
    \chi_+ &= u\chi^\dagger u + u^\dagger \chi u^\dagger, & 
    \hat\chi_+ &= \chi_+ - \frac{1}{2}\Tr{\chi_+} \id,
\end{align}
where $\hat\chi_+$ is the traceless version of $\chi_+$, i.e. the coefficient in front of the trace is equal to the inverse of $\Tr{\id}$. 
The covariant derivative of the nucleon field is given by this expression: 
\begin{align}
    D_\mu N &= \partial_\mu N + \Gamma_\mu N,
\end{align}
where $\Gamma^\mu$ is again the vector current, and finally we include vector and axial-vector sources via: 
\begin{subequations}
\begin{align}
    F^+_{\mu\nu} &= u^\dagger F_{\mu\nu}^R u + u F_{\mu\nu}^L u^\dagger, \\
	F_{\mu\nu}^R &= \partial_\mu r_\nu - \partial_\nu r_\mu -\im [r_\nu,r_\mu],\\
	F_{\mu\nu}^L &= \partial_\mu l_\nu - \partial_\nu l_\mu -\im [l_\nu,l_\mu],
\end{align}
\end{subequations}
where $r_\mu = (v_\mu+a_\mu)/2$ and $\ell_\mu = (v_\mu-a_\mu)/2$.

\section{Interaction Terms}
\label{app:lagrangian}
In this section, we list all the interactions in the Lagrangian that we use in this work. 
The particles that interact in the corresponding vertex are indicated via subscripts. 
First, these are the interactions coming from the chiral Lagrangian in \cref{eq:pionLag}, involving only the pion field and external sources:
\begin{subequations}
\begin{align}
    \mathcal L_{\pi p}^{(2)} &= 2 fB_0 \pi^i p^i, \\
    \mathcal L_{\pi^3 p}^{(2)} &= -\frac{B_0}{5f}p^i \pi^i \pi^j\pi^j, \\
    \mathcal L_{\pi^4}^{(2)} &= \frac{1}{10f^2}\partial_\mu \pi^i \partial^\mu \pi^j \pi^k\pi^l (3\delta^{ik}\delta^{jl}-\delta^{ij}\delta^{kl})\nonumber\\&\qquad -\frac{m B_0}{20f^2}\pi^i\pi^i\pi^j\pi^j, 
\end{align}
\end{subequations}
Using the first term in the expansion of the pion-nucleon interaction operator $A$ in \cref{eq:A1}, we find the following interactions relevant for this work (note that the only parameter is $g_A$, the rest is fixed by chiral symmetry; also additionally to the fields indicated in the subscript, a nucleon enters and exits the vertex):
\begin{subequations}
\begin{align}
    A_{\pi}^{(1)} &= \frac{g_A}{2f}\gamma^\mu\gamma^5 \partial_\mu \pi^i \tau^i, \\
    A_{\pi\pi}^{(1)} &= \frac{1}{4f^2} \gamma^\mu \pi^i \partial_\mu \pi^j \epsilon^{ijk} \tau^k, \\
    A_{\pi^3}^{(1)} &= \frac{g_A}{20f^3} \gamma^\mu\gamma^5 \left[ 3\pi^i \pi^j \partial_\mu\pi^j - \pi^j\pi^j \partial_\mu\pi^i \right]\tau^i.
\end{align}
\end{subequations}
From the second term of the pion-nucleon interaction operator, $A^{(2)}$ in \cref{eq:A2}, we get these interactions:
\begin{subequations}
\begin{align}
    A_{\pi p}^{(2)} &= -\frac{8c_1B_0}{f}p^i\pi^i, \\
    A_{\pi\pi}^{(2)} &= \frac{4B_0 c_1 m}{f^2} \pi^i\pi^i + \frac{c_2}{f^2m_N^2}\partial_\mu \pi^i \partial_\nu \pi^i \partial^\mu\partial^\nu \nonumber\\&\qquad- \frac{c_3}{f^2} \partial_\mu \pi^i \partial^\mu \pi^i + \frac{\im c_4}{f^2}\epsilon^{ijk}\tau^k \gamma^\mu \gamma^\nu \partial_\mu \pi^i \partial_\nu \pi^j,
\end{align}
\end{subequations}
and we also used the fact that the following vertices do not exist: $A_{\pi}^{(2)} = 0$, $A_{\pi^3}^{(2)} = 0$. 

Interactions obtained from the $SU(3)$ Lagrangian, \cref{eq:SU3Lag}, are listed below. 
The interactions without nucleons are: 
\begin{subequations}
\begin{align}
	\mathcal L_{p^3 \pi^0} = \mathcal L_{p^8 \eta_8} &= 2 B_0 f 
\end{align}
\end{subequations}
where we also note that $\mathcal L_{p^0 \pi^0} = \mathcal L_{p^0 \eta} = \mathcal L_{p^8 \pi^0} = 0$. 
The interactions containing nucleons and external currents are given below. 
In the following, the upper sign corresponds to a proton ($\bar NN= \bar pp$) vertex and the lower sign corresponds to a neutron ($\bar NN= \bar nn$) vertex.  
\begin{subequations}
\begin{align}
	\mathcal L_{\bar NN p^8 \pi^0} = \mathcal L_{\bar NN p^3 \eta_8} &= \pm \frac{4 B_0 (b_D+b_F)}{\sqrt{3} f} \\
	\mathcal L_{\bar NN p^8 \eta_8} &= \frac{4  B_0 (6 b_0+5 b_D-3 b_F)}{3 f} \\
	\mathcal L_{\bar NN p^3 \pi^0}&= \frac{4  B_0 (2 b_0+b_D+b_F)}{f} \\
	\mathcal L_{\bar NN p^0 \eta_8} &= -\frac{4 \sqrt{2} B_0 (b_D-3 b_F)}{3 f} \\
	\mathcal L_{\bar NN p^0 \pi^0} &= \pm\frac{4 \sqrt{\frac{2}{3}} B_0 (b_D+b_F)}{f}
\end{align}
\end{subequations}
and the interactions containing no external currents are: 
\begin{widetext}
\begin{subequations}
\begin{align}
	\mathcal L_{\bar NN \pi^0 \eta_8} &= -\frac{4 B_0 \delta m (2 b_0 +b_D+b_F)}{\sqrt{3} f^2} \mp \frac{4 B_0 m (b_D+b_F)}{\sqrt 3 f^2} \label{eq:Lagrangian-breaking1} \\
	\mathcal L_{\bar NN \eta_8\eta_8} &= -\frac{2 B_0 [m (2 b_0+b_D+b_F)+4 m_s (b_0+b_D-b_F))}{3 f^2} \mp \frac{2 B_0 \delta m  (b_D+b_F)}{3 f^2} \label{eq:Lagrangian-breaking2}
\end{align}
\end{subequations}
\end{widetext}
Note that the terms in \cref{eq:Lagrangian-breaking1} break isospin either via non-equal quark masses $\delta m\neq 0$ or via the isospin asymmetry of the nuclear matter (expressed by the different signs in proton and neutron interactions), while \cref{eq:Lagrangian-breaking2} is isospin-even.

\bibliographystyle{apsrev4-2}
\bibliography{references.bib}

\end{document}